\documentclass[aps,prb,twocolumn,groupedaddress,showpacs]{revtex4}

\usepackage{graphicx}
\usepackage{amsmath}
\usepackage{amssymb}
\usepackage{dcolumn}

\begin{document}


\title{Conductance of the single-electron transistor:
       A comparison of experimental data with Monte Carlo calculations}
\author{C. Wallisser}
\email{Christoph.Wallisser@ifp.fzk.de}
\altaffiliation[Also at ]{Fakult\"at f\"ur Physik,
             Universit\"at Karlsruhe,
             D--76128 Karlsruhe, Germany}
\author{B. Limbach}
\altaffiliation[Also at ]{Fakult\"at f\"ur Physik,
             Universit\"at Karlsruhe,
             D--76128 Karlsruhe, Germany}
\author{P. vom Stein}
\altaffiliation[Also at ]{Fakult\"at f\"ur Physik,
             Universit\"at Karlsruhe,
             D--76128 Karlsruhe, Germany}
\author{R. Sch\"afer}
\affiliation{Forschungszentrum Karlsruhe,
             Institut f\"ur Festk\"orperphysik, Postfach 3640,
             D--76121 Karlsruhe, Germany}
\author{C. Theis}
\email{theis@physik.uni-freiburg.de}
\author{G. G\"oppert}
\author{H. Grabert}
\affiliation{Fakult\"at f\"ur Physik,
             Albert-Ludwigs-Universit\"at,
             Hermann-Herder-Stra{\ss}e 3,
             D--79104 Freiburg, Germany}

\date{\today}

\begin{abstract}
We report on experimental results for the conductance of metallic
single-electron transistors as a function of temperature, gate
voltage and dimensionless conductance. In contrast to previous
experiments our transistor layout allows for a direct measurement
of the parallel conductance and no ad hoc assumptions on the
symmetry of the transistors are necessary. Thus we can make a
comparison between our data and theoretical predictions without
any adjustable parameter. Even for rather weakly conducting
transistors significant deviations from the perturbative results
are noted. On the other hand, path integral Monte Carlo
calculations show remarkable agreement with experiments for the
whole range of temperatures and conductances.
\end{abstract}


\maketitle

\newcommand*{\Usd}{\ensuremath{U_{\text{sd}}}}
\newcommand*{\Isd}{\ensuremath{I_{\text{sd}}}}

\section{\label{sec:intro}Introduction}

The usual single-electron transistor (SET) layout
\cite{joyez97,chouv99} consists of two ultrasmall tunnel
junctions with conductances $G_{s,d}$ and capacitances $C_{s,d}$,
respectively. The junctions in series are biased by a voltage
$\Usd$ and the island between the tunneling barriers is coupled
to a gate voltage $U_g$ via a capacitance $C_g$.

This setup has been used as the basic device to study the Coulomb
blockade \cite{grabe92}. Its transport properties are governed by
two dimensionless parameters. Firstly, the dimensionless inverse
temperature $\beta E_c$ relates $\beta=(k_B T)^{-1}$ to the
charging energy $E_c=e^2/(2 C_\Sigma)$ with
$C_\Sigma=C_s+C_d+C_g$ being the total island capacitance. This
parameter determines how far the electrostatic blockade of the
source-drain current is lifted by thermal excitations. Secondly,
the dimensionless parallel conductance $g=(G_s+G_d)/G_K$,
$G_K=e^2/h$ being the conductance quantum, measures how much the
quantization of the island charge is smeared by quantum
fluctuations. The charge on the island can be controlled by means
of the gate voltage $U_g$. The linear-response conductance $G$
between source and drain is a periodic function of $U_g$ taking
on values between $G_{\text{min}}$ and $G_{\text{max}}$. Since the
control of a device by Coulomb blockade exploits a large
difference between $G_{\text{min}}$ and $G_{\text{max}}$ a detailed
understanding of the washout of Coulomb blockade effects by
thermal and quantum fluctuations is crucial for the optimal
design of fast and reliable Coulomb blockade devices.

Theoretical work has mainly focused on the limits of small $g$ or
$\beta E_c$, respectively. In the weakly conducting regime, $g
\ll 1$, perturbation theory \cite{koeni97,koeni98a,schoe94}
should be sufficient to describe the experimental data. For $\beta
E_c \ll 1$ it is more suitable to formulate the problem in terms
of a path integral which may be evaluated semiclassically
\cite{golub96,goepp98,goepp00a}. Only recently the use of path
integral Monte Carlo (PIMC) techniques was proposed to calculate
the conductance of the SET over the whole range of experimentally
accessible parameters \cite{goepp00b}. Especially the regime of
large parallel conductance $g$ is of interest for technological
applications \cite{lafar91,clela92,pekol94}.

This recent work has pointed to a shortcoming of previous
experiments. The first step of a comparison between theory and
experiment is the determination of the parameters $g$ and $E_c$.
For the SET layout described above one can obtain only the series
conductance $G_\Sigma$ of the two junctions which does not
suffice to calculate the dimensionless parallel conductance $g$
without further assumption. \citeauthor{joyez97}\cite{joyez97}
assumed that their SETs were symmetric, i.\,e.\ built up of two
identical tunnel barriers which turned out to be a good
approximation for the first three samples measured in the
experiment. However, comparing the data for the high conductance
sample of Ref.~\onlinecite{joyez97} to their PIMC simulation
results, \citeauthor{goepp00b} \cite{goepp00b} found strong
indications that the tunnel barriers could be asymmetric.

\begin{figure}
  
  \includegraphics[width=\linewidth]{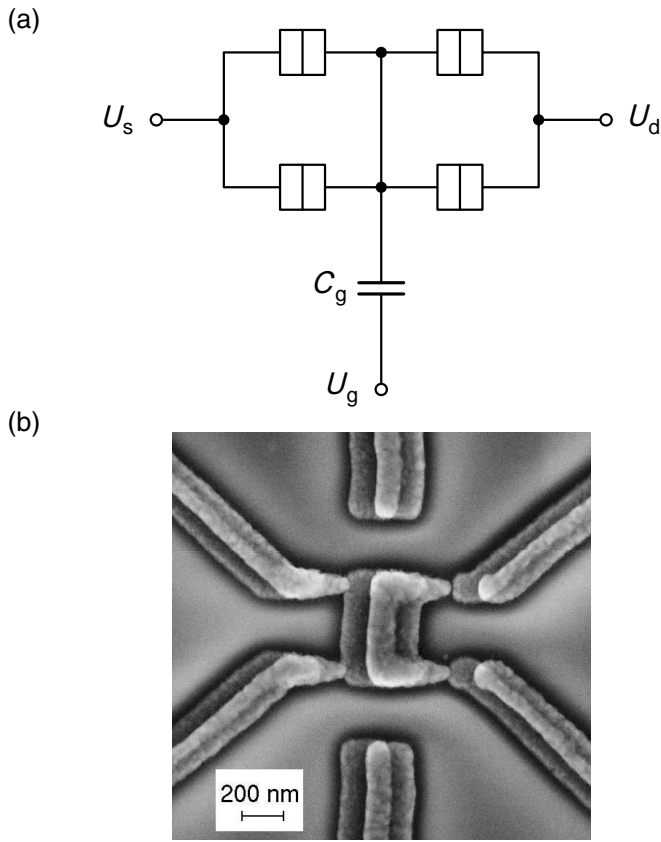}

  \caption{a) Circuit diagram for the four junction layout when
    operated as SET. b) SEM picture of a four junction SET. The layout
    contains two symmetrically arranged gate fingers to avoid asymmetrical
    proximity effects in the lithography process. In the experiments both
    gates are connected in parallel.\label{fig:sem}}
\end{figure}

Since an unambiguous determination of the relevant parameters is a
prerequisite of a thorough comparison between theoretical and experimental
results we have developed a transistor design with four tunnel junctions
connected to the island (see Fig.~\ref{fig:sem}). This arrangement allows the
determination of the individual resistances of each junction and therefore the
direct measurement of the tunneling strength parameter $g$. It is operated
as SET by connecting two of the junctions in parallel to form source and
drain, respectively.

In Sec.~\ref{sec:experiment} we present experimental details
about the sample fabrication, the determination of the parameters
and the conductance measurements. In Sec.~\ref{sec:theory} we
give the path integral formulation which is used for the imaginary
time quantum Monte Carlo simulation. We summarize how the
conductance can be calculated from the simulation data by use of
the singular value decomposition (SVD) analytical continuation
scheme. In Sec.~\ref{sec:results} we compare experimental and
theoretical results. At low conductance the perturbation theory
is sufficient to describe the temperature dependence of
$G_{\text{max}}$, but for $G_{\text{min}}$ significant deviations are
observed. At high conductance, we make a direct comparison of the
experimental data with path integral Monte Carlo results.

\section{Experiment\label{sec:experiment}}

Seven samples with varying tunneling strengths were investigated. The
samples were fabricated from aluminum by standard e-beam lithography in
combination with two-angle shadow evaporation. The evaporation is performed
by electrical heating a tungsten wire which holds a drop of aluminum. Tunnel
barriers of different strengths could be produced by a variation of the
oxygen pressure applied to the evaporation chamber between the two
evaporation steps.

We used two different layouts: firstly, the usual SET design with two
junctions, forming a small island in between, and a straight gate finger
from the side pointing towards the island, secondly, the design which is
shown on the SEM picture in Fig.~\ref{fig:sem}. In this design one can
determine the individual tunnel resistances by measuring the current in
response to an applied voltage bias across different combinations of the
four tunnel junctions. The measurements presented in this article are
performed on four standard SETs (samples I--IV), one four-contact sample
(sample VII) and two samples which also have the four-contact layout, but
turned out to have only three working tunnel junctions (samples V and VI).
Nevertheless, this is sufficient to determine all individual tunnel
resistances. The sample parameters are given in Table \ref{tab:samples}.
\begin{table*}
\begin{ruledtabular}
\begin{tabular}{ldcddddddd}
 &
 \multicolumn{1}{c}{$G_{\text{cl}}^{-1}$} &
 \multicolumn{1}{c}{$C_\Sigma$}  &
 \multicolumn{1}{c}{$C_g$} &
 \multicolumn{1}{c}{$G_s^{-1}$} &
 \multicolumn{1}{c}{$G_{s'}^{-1}$} &
 \multicolumn{1}{c}{$G_d^{-1}$} &
 \multicolumn{1}{c}{$G_{d'}^{-1}$} &
 \multicolumn{1}{c}{$E_c$} &
 \multicolumn{1}{c}{$g$} \\
 &
 \multicolumn{1}{c}{(k$\Omega$)} &
 \multicolumn{1}{c}{(aF)} &
 \multicolumn{1}{c}{(aF)}&
 \multicolumn{1}{c}{(k$\Omega$)}&
 \multicolumn{1}{c}{(k$\Omega$)}&
 \multicolumn{1}{c}{(k$\Omega$)}&
 \multicolumn{1}{c}{($k_B$\,K)}&
 \multicolumn{1}{c}{($k_B$\,K)} & \\
\hline
 I & 128.0 & 220 & 25.6 & & & & & 4.25 & 0.80 \ (1.10)\\
 II & 74.7 & 240 & 27.6 & & & & & 3.87 & 1.39 \\
 III & 19.2 & 250 & 27.5 & & & & & 3.70 & 5.38 \\
 IV & 17.2 & 280 & 30.2 & & & & & 3.30 & 5.98 \\
 V & 74.5 & 278 & 18.0 & 39.4 & & 62.5 & 79.7 & 3.4 & 1.40 \\
 VI & 59.4 & 344 & 18.0 & 22.7 & & 104.2 &56.1& 2.7 & 1.85 \\
 VII & 23.0 & 497 & 19.0 &  20.3& 16.4 & 31.7 & 23.8 & 1.87 & 4.75 \\
\end{tabular}
\end{ruledtabular}
\caption{Parameters of the samples I--VII. $G_{\text{cl}}$ denotes
the high temperature conductance of the SET. The gate capacitance
$C_g$ is extracted from the period of the Coulomb oscillations
within an accuracy of about 1\%. Samples V-VII consist of at
least three tunnel junctions. The individual conductances
$G_{s,s'}$ and $G_{d,d'}$ for these samples were evaluated by
simple algebra from values of $G_\Sigma$, which had been measured
at different pairs of tunnel junctions as described in the text.
For the 2-junction samples I--IV, the given value of $g$ equals
$(G_s+G_d)/(2G_K)$, an expression valid for symmetric SETs only.
For the asymmetric sample I, the value derived from a comparison with the
second-order perturbation theory is given in brackets (see
Fig.~\ref{fig:2con}).
\label{tab:samples}}
\end{table*}
The measurements are performed in a top-loading dilution refrigerator in the
temperature range from 25~mK to 18~K. The samples are mounted within a
well-shielded metallic cavity. All electrical wiring into the cavity is made
of highly resistive leads (32~$\Omega$/m, diameter 0.23~mm) which are fed
through stainless steel capillaries with an inner diameter of 0.34~mm and a
length of 1~m. The capillaries are wound up in a compact coil which is held
in thermal equilibrium with the sample. These feedthroughs form resistive
co-axial cables. They provide a damping exceeding 200~dB in the frequency
range from 20~GHz to 6~THz, as calculated by classical electrodynamics
taking the skin effect into account \cite{zorin95}. The validity of the used
formulae was checked experimentally in the frequency range up to 20~GHz
using a spectrum analyzer. Additional rf-filtering is performed at room
temperature at the entrance to the cryostat.

To measure the conductance of the transistors they are biased with a voltage
and the resulting current is measured with an operational amplifier at the
top of the cryostat. The resolution of the current measurement is better
than 100~fA. To gain resolution and to circumvent the 1/f-increase of noise
at low frequencies an AC component of $\approx 10$~Hz is added to the
biasing DC voltage simultaneously and the resulting AC component of the
current is measured with a lock-in amplifier. The DC measurement is used to
ensure that the measurement is performed at vanishing bias and stays in the
linear-response regime.

To determine the conductance $G_\Sigma$ of two tunnel junctions in series,
we measure the $\Isd\Usd$ characteristic up to a bias voltage of maximal
$\pm$20~mV. We define the asymptotic slope at large bias voltages as
$G_\Sigma$. The four-junction design allows six different configurations
of two tunnel junctions connected in series. From the corresponding set
$G_{\Sigma,i}, \; i=1,...,6$, the individual tunnel junction conductances
can be derived with simple algebra. Their values are given in table
\ref{tab:samples}. Thus this layout enables us to actually measure the
coupling strength parameter $g$ directly. The SET investigated in the
following experiments was formed by using all four tunnel junctions, where
source and drain were made by connecting two tunnel junctions in parallel,
respectively (see Fig.~\ref{fig:sem}).

\section{Theory\label{sec:theory}}

\subsection{Path integral formulation\label{subsec:pathintegral}}

The linear DC conductance $G$ represents a transport coefficient
which can be expressed by correlation functions of the system
using a Kubo formula \cite{goepp00b}. Defining the current through
the SET as the average of the current through the source and drain
junctions $I=(I_s+I_d)/2$ the conductance $G$ may be connected to
the spectrum of the current autocorrelation function $F(t)=
\langle I(t) I(0) \rangle$, i.e.
\begin{equation}
G=\frac{\beta}{2}\tilde{F}(\omega=0) \label{eq:kubo}
\end{equation}
with $\tilde{F}(\omega)$ denoting the Fourier transform of $F(t)$.
The calculation of the current correlator may be done in the phase
representation \cite{schoe90}, i.e. in terms of the phase
variable $\varphi$ which is conjugate to the charge $q$ on the
island. For imaginary time $\tau$ one gets \cite{goepp98}
\begin{equation}
F(\tau)= 4 \pi G_{\text{cl}} \alpha(\tau) C(\tau) \label{eq:Iicf}
\end{equation}
with the classical high temperature conductance $G_{\text{cl}}$
and
\begin{subequations}
  \label{eq:alpha}
  \begin{eqnarray}
    \alpha(\tau)&=&\frac{1}{2 \pi}
    \int_{-\infty}^\infty d\omega \; \tilde{\alpha}(\omega) \;
    e^{-\tau \omega}, \label{eq:alphaa} \\
    \tilde{\alpha}(\omega)&=&\frac{\hbar}{2 \pi}
    \frac{\omega}{1-e^{-\hbar \beta \omega}}. \label{eq:alphab}
  \end{eqnarray}
\end{subequations}
The cosine correlation function $C(\tau)= \langle \cos \left[
\varphi(\tau)-\varphi(0)\right] \rangle$ has the formally exact
path integral representation \cite{goepp98}

\begin{eqnarray}
C(\tau)= \frac{1}{Z} \sum_{k=-\infty}^\infty
\int\limits_{\varphi(0)=0}^{\varphi(\beta \hbar)=2 \pi k}
\!\!\!\!\!\! && {\cal D}\varphi \; e^{-\frac{1}{\hbar}S[\varphi]+2
\pi i k n_g} \; \times \nonumber
\\ \times && \cos\left[\varphi(\tau)-\varphi(0)\right]
\label{eq:pathintegral}
\end{eqnarray}
%
with the partition function
\begin{equation}
Z=\sum_k \int {\cal D} \varphi \; e^{-\frac{1}{\hbar} S[\varphi]}
e^{2 \pi i k n_g} \label{eq:Z}
\end{equation}
and the dimensionless gate voltage $n_g=(U_g C_g)/e$. The
Euclidian action $S[\varphi]=S_C[\varphi]+S_T[\varphi]$ splits
into the Coulomb action
\begin{equation}
S_C[\varphi]=\int_0^{\beta \hbar} d\tau \frac{\hbar^2
\dot{\varphi}^2(\tau)}{4 E_c} \label{eq:Sc}
\end{equation}
which describes the charging of the island and the tunneling action
\begin{equation}
S_T[\varphi]=2 g \int\limits_0^{\beta \hbar} d\tau
\int\limits_0^{\beta \hbar} d\tau' \; \alpha(\tau-\tau') \;
\sin^2\left[\frac{\varphi(\tau)-\varphi(\tau')}{2}\right]
\label{eq:St}
\end{equation}
that expresses the influence of tunneling processes on the
dynamics of the phase variable $\varphi$.

Equation (\ref{eq:pathintegral}) may serve as the starting point
for analytical calculations \cite{goepp98,goepp00a} or numerical
work \cite{goepp00b}. In the latter case the interval $[0,\beta
\hbar]$ is divided into $N$ Trotter slices and the
multidimensional integral is calculated using Monte Carlo methods
\cite{binde98,suzuk93}. Since the action $S[\varphi]$ is real, the
Metropolis algorithm can be applied for an importance sampling of
the configurations $\{ \varphi_i \equiv \varphi (\tau_i) |
i=0,...,N \}$ and the winding number $k$.

We did Monte Carlo simulations for fixed tunneling strength
$g=4.75$ over a range of inverse temperatures $\beta E_c \in [0.5,
21.0]$ which will be compared to our experimental findings in
Section \ref{sec:results}. For each temperature the system was
equilibrated during several million Monte Carlo steps.
Measurements were then carried out for another 5--32 million
sweeps depending on the value of $\beta E_c$. Especially for low
temperatures it is necessary to increase the number of
measurements to get reliable statistics of the data. We
ensured that the error of the correlation function is always less
than $3\%$. Over the whole range of temperatures we chose
$N=200$ Trotter slices fulfilling the convergence criterion
$N\ge5 \beta E_c$ used in earlier work \cite{goepp00b,herre99}.
The conductance was calculated for $200$ values of the
dimensionless gate voltage spanning the range $n_g\in [0,0.5]$.

Additionally we examined single electron transistors with
different parallel conductances $g \in [2.0,15.0]$ for fixed
temperature $1/(\beta E_c)=0.05$. Here we chose $N=250$
Trotter slices and did up to 23 million measurement sweeps. We
found that the convergence was slower for small values of the
dimensionless conductance.

\subsection{Analytic continuation\label{subsec:svd}}

Having calculated the cosine correlation function $C(\tau)$ using
the PIMC method we still have to solve Eqs.~(\ref{eq:kubo}) and
(\ref{eq:Iicf}) for the linear conductance $G$. With the positive
and symmetric spectral function \cite{goepp00b}
$A(\omega)=\tilde{C}(\omega) (1-e^{-\beta \hbar \omega})/\omega$
one can write the Fourier transformation of $C(\tau)$ in
imaginary time as
\begin{equation}
C=KA \label{eq:ip}
\end{equation}
where the integral operator $K$ is given by
\begin{equation}
(KA)(\tau)= \frac{1}{2 \pi} \int_0^\infty d\omega \; \frac{\omega
\cosh\left[(\frac{\beta \hbar}{2}-\tau)
\omega\right]}{\sinh(\frac{\beta \hbar}{2} \omega)} A(\omega)
\label{eq:intop}
\end{equation}
and Eqs.~(\ref{eq:kubo}) and (\ref{eq:Iicf}) can be combined to
give
\begin{equation}
G=\frac{\beta \hbar G_{\text{cl}}}{2 \pi} \int_0^\infty d\omega
\; \frac{\omega^2}{\cosh(\beta \hbar \omega)-1} A(\omega)\;.
\label{eq:cond}
\end{equation}
Before we can calculate $G$ using Eq.~(\ref{eq:cond}) it is
necessary to invert the integral transform of Eq.~(\ref{eq:ip}) to
obtain $A(\omega)$. This problem is similar to the analytical
continuation of a function from the imaginary to the real axis
\cite{galli98}. It belongs to the class of \textit{ill-posed}
inverse problems. A straightforward inversion of Eq.~(\ref{eq:ip})
fails, because the integral operator $K$ is almost singular and
the statistical error in the data $C(\tau)$ is strongly amplified,
making the result $A(\omega)$ meaningless. To deal with this
problem, we use a recently developed method \cite{devil99} based,
upon the singular value decomposition of the integral operator.

From the SVD we get the singular system of $K$ which fulfills
\begin{subequations}
  \label{eq:singularsystem}
  \begin{eqnarray}
    (K u_j)(\omega) &=& \sigma_j v_j(\omega) \label{eq:rsv} \\
    (K^\dagger v_j)(\tau) &=& \sigma_j u_j(\tau) \label{eq:lsv} \\
    \sigma_0 \ge \sigma_1 &\ge& \sigma_2 \ge ... \ge 0\;.  \label{eq:sv}
  \end{eqnarray}
\end{subequations}
The functions $u_j(\tau)$ and $v_j(\omega)$ are called the right
and left singular vectors, respectively. The real and positive
$\sigma_j$ are the singular values of $K$. The formal solution to
the inverse problem (\ref{eq:ip}) is given as \cite{devil99}
\begin{equation}
A(\omega)=\sum_{j=0}^\infty \frac{c_j}{\sigma_j} v_j(\omega)
\label{eq:formalsolution}
\end{equation}
with the coefficients
\begin{equation}
c_j=\int_0^{\beta\hbar} d\tau \; C(\tau) u_j(\tau).
\label{eq:expansioncoefficients}
\end{equation}

Since the numerical calculation of the correlation function
$C(\tau)$ can only be done for a discrete set of points and
additionally introduces a statistical error, the expansion
coefficients (\ref{eq:expansioncoefficients}) have a limited
accuracy. Taking into account that for an ill-posed problem such
as (\ref{eq:ip}) the singular values $\sigma_j$ vanish
overexponentially with increasing $j$ it becomes obvious that only
the first few terms in the expansion (\ref{eq:formalsolution})
contain meaningful information whereas the higher terms will just
corrupt the result by amplifying the noise of the data. This idea
is used in the truncated SVD approach \cite{galli98,devil99}
which truncates the summation in (\ref{eq:formalsolution}) by
neglecting all terms for which $\sigma_j/\sigma_0$ is smaller
than the statistical error of the correlation function $C(\tau)$.

Apart from Eq.~(\ref{eq:ip}) we still have supplementary
information, namely that the spectral function $A(\omega)$ is
positive and symmetric. To ensure positiveness one may use a
"triangular window" \cite{berte88}, i.e. in the truncated
singular value decomposition one multiplies the singular values
with a weight factor which falls of linearly from one to zero.
Previous studies \cite{galli98, berte88} have shown that this
implementation of positiveness reduces the resolution of the
method. Recent work \cite{devil99} has shown how to use
supplementary information of positiveness to enhance the resolution
of the singular value decomposition method. The idea is to
determine additional expansion coefficients which cannot be
inferred from the inverse problem. They can be fixed by the
constraints that the result shall be positive and the difference
to the truncated SVD solution shall be minimal. Further details
about the implementation of the method can be found in
Ref.~\onlinecite{devil99}.

The approach used in this work is limited only by the statistical
error in the correlation function (\ref{eq:pathintegral}). The
error of our PIMC calculation and the subsequent analytical
continuation was estimated as follows. First we determined the
statistical error of the Monte Carlo data. Then we produced an
ensemble of data sets with different realizations of the error by
adding Gaussian distributed random noise of the given size. The
error bars shown in Figs.~\ref{fig:4conminmax}, \ref{fig:condbeta}
and \ref{fig:condalpha} were then calculated from the maximum and
minimum result of the analytical continuation.

\section{Results and Discussion\label{sec:results}}

In this section we present experimental data and theoretical
results for the minimum and maximum linear response conductance,
$G_{\text{min}}$ and $G_{\text{max}}$, respectively, as a
function of the dimensionless parallel conductance $g$ and the
temperature. Before we turn to the comparison between
experimental and theoretical data we would like to summarize our
findings about the determination of the experimental parameters.

\begin{figure}
  \includegraphics[width=\linewidth]{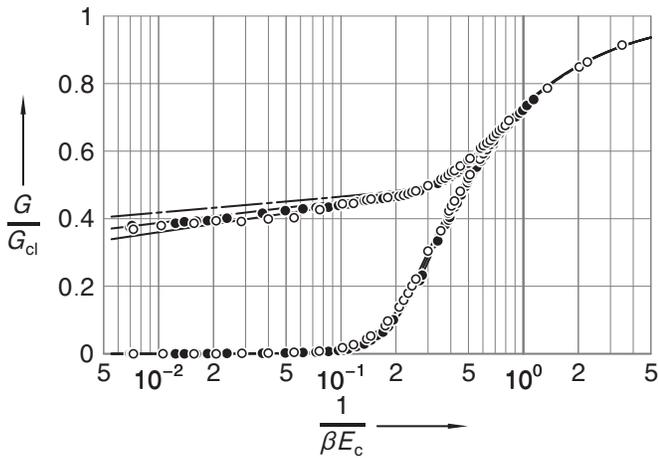}
  \caption{$G_{\text{max}}$ and $G_{\text{min}}$ normalized to
    the high temperature conductance $G_{\text{cl}}$ for sample I
    ($\bullet$) and II ($\circ$) as a function of the normalized
    temperature. $G_{\text{max}}$ and $G_{\text{min}}$ are the maximum and
    minimum linear response conductance observed as a function of the gate
    voltage $U_g$, respectively.  Lines correspond to the predictions of a
    second-order perturbative expansion in $g$ (solid line: $g=1.4$, dashed
    line: $g=1.1$, dashed-dotted line: $g=0.8$).  The data are best
    described by $g=1.1$ ($\bullet$) and $g=1.4$ ($\circ$).
    \label{fig:2con}}
\end{figure}

The four-junction layout allows us to determine the dimensionless parallel
conductance unambiguously offering the possibility of a direct comparison
between experiment and theory. On the other hand measurements carried out on
the two-junction samples I and II (see Fig.~\ref{fig:2con}) clearly
demonstrate a problem also encountered in Ref.~\onlinecite{goepp00b}. The
assumption that the conductance is distributed equally among both tunnel
junctions would lead to $g_{\text{I}}=0.80$ and $g_{\text{II}}=1.39$ for the
samples I and II in contradiction to their almost identical temperature
dependence of both $G_{\text{min}}$ and $G_{\text{max}}$. Since the maximum
conductance $G_{\text{max}}$ in the weak tunneling regime should fall off
proportional to $g \ln (\beta E_c/\pi)$ for low temperatures \cite{koeni98a}
Fig.~\ref{fig:2con} shows that the correct parallel conductances of those
samples are almost equal.
Our four-junction samples allow us to check this assumption directly (see
below). We found that the tunnel junctions are \textit{not identical}
although they are produced simultaneously during shadow evaporation. This is
not surprising as we tried to produce contacts with oxide barriers as thin
as possible and very small junction areas. At the borderline between
functioning and broken contacts the unavoidable variations in the
fabrication procedure become visible as large fluctuations in the
conductances of different junctions.

\begin{figure}
\includegraphics[width=\linewidth]{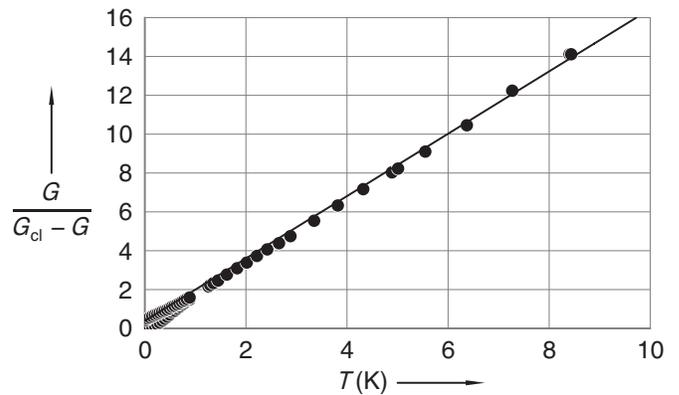}
\caption{Inverse conductance reduction $G/(G_{\text{cl}}-G)$ as
  a function of temperature. ($\bullet$): Experimental data for sample VII,
  measured at vanishing source-drain voltage. (\textbf{---}):
  High-temperature expansion Eq.~(\ref{eq:hte}) with $E_c$ chosen so as to
  fit the data.
  \label{fig:semifit}}
\end{figure}

In previous experiments \cite{joyez97,chouv99,lu98} several methods for the
determination of the charging energy $E_c$ were proposed.  As already
mentioned in Ref.~\onlinecite{chouv99} the determination of the charging
energy from the offset of the $\Isd\Usd$ curve is not very accurate. For
sample V and VII we have analyzed the subgap resonances observed in the SET
in the superconducting state to obtain the renormalized charging energy
$E_c^S$ as described in Ref.~\onlinecite{joyez97}. \citeauthor{joyez97} also
give a perturbative result which connects $E_c^S$ to the bare charging
energy $E_c$ \footnote{The formula given in Ref.~\onlinecite{joyez97}
  contains a typo.  Instead of $x/\pi$ the prefactor in the definition of
  $f(x)$ should read $x/\pi^2$.}.  Unfortunately it is only valid up to
$\mathcal{O}(g^2)$. In our experiments we found that this is not sufficient
for a description of high conductance single-electron transistors. In our
opinion the best method for the determination of the charging energy is a
comparison of the high-temperature experimental data to semiclassical
calculations \cite{goepp98}. Apart from the full semiclassical expressions
we tested the high-temperature expansion
\begin{equation}
   \frac{G}{G_{\text{cl}}-G}=
      \frac{3k_BT}{E_c}+\frac{27g\zeta(3)}{2\pi^4}-\frac{2}{5}
   \label{eq:hte}
\end{equation}
using only data with $k_B T \gg g E_c/ 2\pi^4$ (see
Fig.~\ref{fig:semifit}). At a dimensionless conductance of
$g=4.75$ the difference in the charging energy between the full
semiclassical result and Eq.~(\ref{eq:hte}) was less than $1\%$.
In contrast to earlier work \cite{chouv99} sufficient experimental
data at high temperatures are available for a determination of
$E_c$. Moreover, Eq.~(\ref{eq:hte}) provides a consistency check for the
dimensionless parallel conductance $g$ which we have determined
independently.

\begin{figure*}
   \parbox[t]{0.01\linewidth}{(a)}
   \parbox[t]{0.46\linewidth}
  {\vspace{0pt}\includegraphics[width=\linewidth]{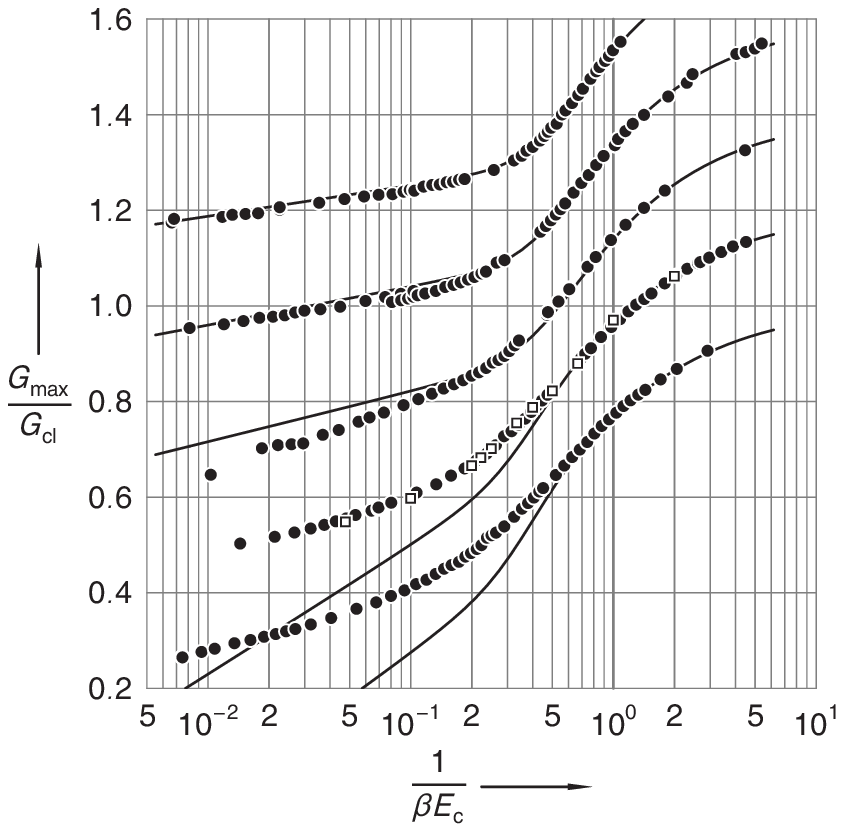}}
   \hfill
   \parbox[t]{0.01\linewidth}{(b)}
   \parbox[t]{0.46\linewidth}
  {\vspace{0pt}\includegraphics[width=\linewidth]{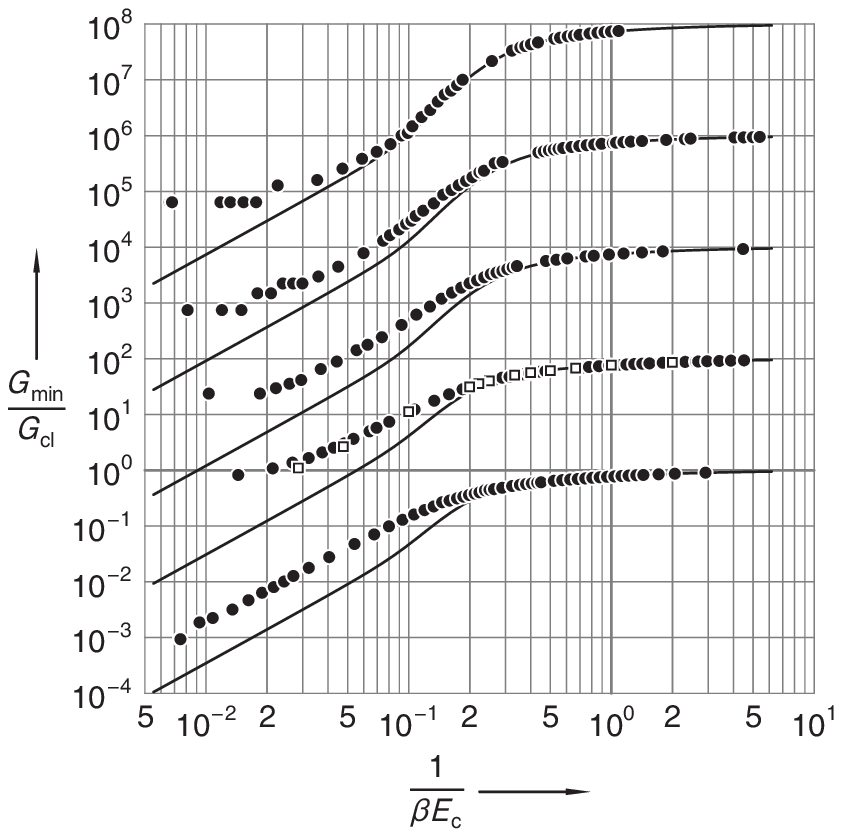}}

  \caption{Maximum (a) and minimum (b) conductance normalized to
    the high temperature conductance $G_{\text{cl}}$ for samples I, V, VI,
    VII, and III with (from top to bottom) $g=1.10, 1.40,1.85, 4.75$, and
    $5.38$ as a function of the normalized temperature. Together with the
    experimental data ($\bullet$) the predictions of the perturbation theory
    in second order (\textbf{---}) are shown for all samples. For sample VII
    the results of the Monte Carlo calculations ($\scriptstyle \square$) are
    given additionally. For the sake of clarity the curves are plotted in
    (a) with a vertical offset increasing by $0.2$ from curve to curve, in
    (b) the different datasets are multiplied by $10^2$, $10^4$, etc.
  \label{fig:4conminmax}}
\end{figure*}

Having determined the experimental parameters we can turn to the comparison
with the theoretical results. First of all we want to compare the data
measured on sample I, III and V--VII with perturbation theory in second
order in the parallel conductance $g$. Fig.~\ref{fig:4conminmax} shows the
maximum and minimum conductance of the single-electron transistor as a
function of temperature. For sample I and V with their moderate $g=1.1$ and
$g=1.40$, respectively, good agreement can be stated for $G_{\text{max}}$.
At higher $g$ (sample III, VI and VII) deviations of increasing size are
visible. Such deviations are not surprising as the perturbation expansion is
not justified in this parameter range. For the minimum conductance
$G_{\text{min}}$ we get deviations from the second-order result for all
values of $g$ investigated. At the lowest temperatures the determination of
the minimum conductance $G_{\text{min}}$ is limited by the resolution of the
lock-in signal which is of the order $10^{-3} G_0$. However the deviations
are also prominent at $1/(\beta E_c) = 0.05$ where the lock-in signal has a
sufficient accuracy. The discrepancy between perturbation theory and our
data increases with $g$, indicating that higher-order corrections have to be
included. With third-order perturbation theory \cite{schoe00} deviations for
$G_{\text{min}}$ are reduced but still visible.

Besides renormalization group methods \cite{koeni98b} the quantum Monte
Carlo approach is the only method which can cover the whole range of
parameters that is accessible in the experiment. For sample VII ($g=4.75$),
which is beyond the perturbative regime, we perform a detailed comparison.

\begin{figure}
  \includegraphics[width=\linewidth]{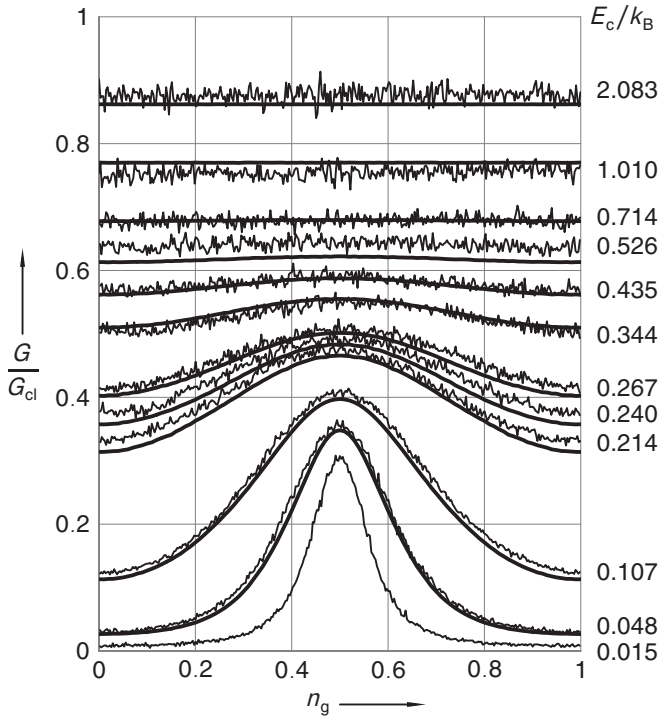}
  \caption{Coulomb oscillations for sample VII ($g=4.75$, thin
    lines) compared with Quantum Monte Carlo calculations (thick lines). The
    conductance $G$ normalized to the high temperature conductance
    $G_{\text{cl}}$ is shown as a function of the dimensionless gate voltage
    $n_g=(C_g U_g)/e$ for the temperatures as given in the right margin. The
    calculations were done at temperatures (from bottom to top) 
    0.048, 0.1, 0.2, 0.22, 0.25, 0.33, 0.4, 0.5, 0.67, 1.0 and 2.0 $E_c/k_B$.
  \label{fig:coszi}}
\end{figure}

In Fig.~\ref{fig:coszi} we show the gate voltage dependent conductance for
inverse temperatures ranging from $1/(\beta E_c)= 0.048$ to $1/(\beta E_c)
=2.0$. We find that the experimental Coulomb oscillations are very well
described by the Monte Carlo calculations. Minor discrepancies at some
temperatures can be attributed to the fact that the temperatures used in the
simulations do not exactly match those of the experimental data.  For the
lowest temperatures of the experiment it was not possible to get converged
Monte Carlo results for the whole range of gate voltages in reasonable time.

\begin{figure}
  \includegraphics[width=\linewidth]{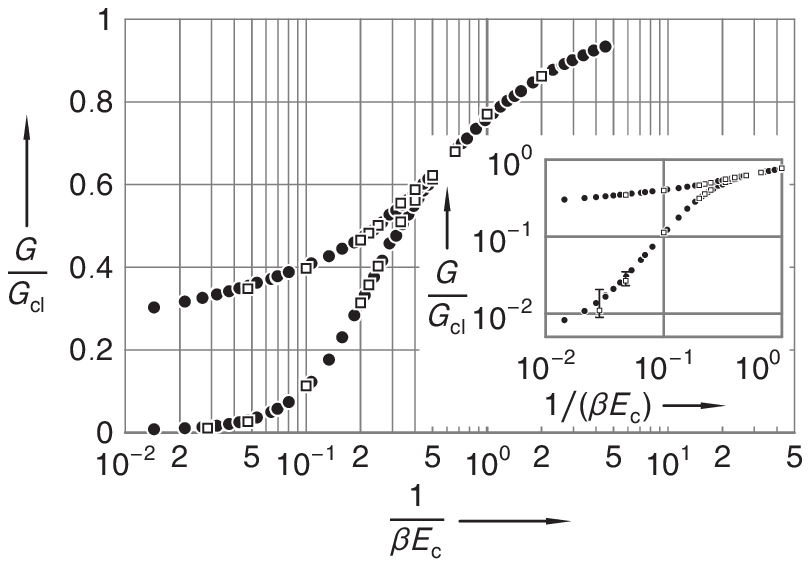}
  \caption{Maximum and minimum linear response conductance
    $G_{\text{min}}$ and $G_{\text{max}}$ for sample VII ($g=4.75$) normalized
    to the high temperature conductance $G_{\text{cl}}$.  The experimental
    data ($\bullet$) are compared to the results of our PIMC simulation
    ($\scriptstyle \square$). Data points for $G_{\text{max}}$ at the lowest
    temperatures have been omitted (see text). The inset shows the same data
    on a logarithmic scale for better comparison of $G_{\text{min}}$.
    Errorbars are only shown if they exceed the symbol size.
  \label{fig:condbeta}}
\end{figure}

In Fig.~\ref{fig:condbeta} the data are analyzed in terms of the minimum and
maximum conductance. Once again the accordance is remarkable. For high
$\beta E_c$ the maximum conductance could not be determined by our Monte
Carlo calculations. Here a limitation of the Monte Carlo procedure becomes
obvious. For low temperatures more terms of the winding number summations in
Eqs.~(\ref{eq:pathintegral}) and (\ref{eq:Z}) are relevant leading to phase
cancellations due to the factors $\exp(2\pi i k n_g)$ which are especially
strong at $n_g=0.5$, i.\,e.\ for the maximum conductance. Thus the
convergence of the Monte Carlo procedure gets slower with decreasing
temperature and the data cannot be determined as accurately.  Since the
analytic continuation is sensible to the statistical error of the data,
reliable results for $G_{\text{max}}$ could not be obtained with reasonable
effort and the data points for $G_{\text{max}}$ at the lowest temperatures
have been omitted. This is not the case for the minimum conductance
$G_{\text{min}}$ as can be seen in the inset of Fig.~\ref{fig:condbeta}.
Here no phase cancellations occur and the experimental and theoretical data
match nicely even on a logarithmic plot. We can also observe that in
contrast to the perturbation theory in second order (cf.\@
Fig.~\ref{fig:4conminmax}), the Monte Carlo approach gives an accurate
description of $G_{\text{min}}$ at low temperatures.

\begin{figure}
  \includegraphics[width=\linewidth]{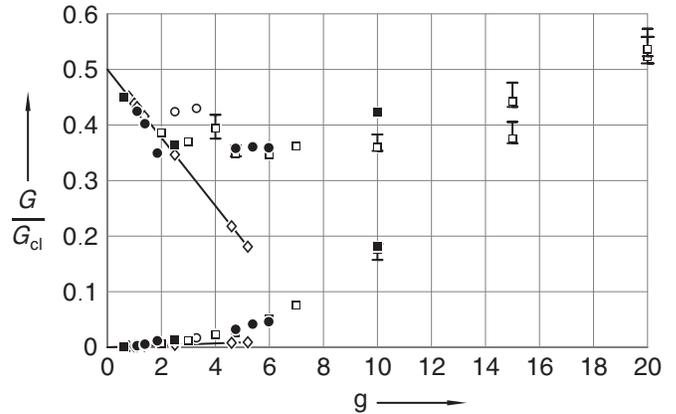}
  \caption{Maximum and minimum conductance for $1/(\beta E_c)
    =0.05$ as a function of the tunneling strength $g$ for all examined
    samples ($\bullet$) in comparison with PIMC calculations ($\scriptstyle
    \square$). Included are also the experimental data of
    \citeauthor{joyez97} \cite{joyez97} ($\scriptstyle \blacksquare$) and
    the results of perturbation theory in second order
    \cite{koeni97,koeni98a} ($\scriptstyle \lozenge$) and third order
    \cite{schoe00} ($\circ$). Errorbars are only shown if they exceed
     the symbolsize.
    \label{fig:condalpha}}
\end{figure}

Finally we have examined the maximum and minimum conductance of
single-electron transistors for varying tunneling strength $g$ at a fixed
temperature $1/(\beta E_c) = 0.05$. The results are shown in
Fig.~\ref{fig:condalpha}. Besides our experimental data the results of
\citeauthor{joyez97} \cite{joyez97} are shown. 
According to Ref.~\onlinecite{goepp00b} the data of
\citeauthor{joyez97} at $g=7.5$ has been displayed at $g=10$. The Monte
Carlo data include also the earlier results of \citeauthor{goepp00b}
\cite{goepp00b}. Once again we observe a reasonable accordance between
theory and experiment. For $g>8$ the comparison is hampered by uncertainties
of the experimental parameters $g$ and $E_c$. The parallel conductance of
earlier experiments was systematically underestimated by the assumption of
symmetry of the SET while for the determination of the charging energy $E_c$
sufficient high temperature data were missing.

Also shown are the predictions of the perturbation expansion in
second\cite{koeni97,koeni98a} and third order\cite{schoe00} in
$g$. The range of validity of the second-order perturbative
approach is limited to $g \le 2.5$ where the maximum conductance
$G_{\text{max}}$ drops with increasing $g$. The plateau and the
following increase of the maximum conductance can not be
described by perturbation theory. Also for $G_{\text{min}}$
deviations occur at $g > 2.5$ while the Monte Carlo approach
gives excellent results up to $g=10$.

\section{\label{sec:conclusion}Conclusion}

We have presented experimental results for the conductance of
single-electron transistors as a function of temperature and
dimensionless gate voltage. The employed four-junction layout for
the SET allows for an unambiguous determination of the physical
parameters $g$ and $E_c$. Thus we were able to clarify and
eliminate problems encountered in earlier experiments. In
particular, comparison with theory can be made without any
adjustable parameter. We have compared the experimental findings
with perturbation theory in second order in $g$ and with results
of PIMC simulations.

Comparison with perturbation theory was made for the maximum and minimum
conductance of five SETs with different tunneling strength. At $g<1.85$ we
found good agreement for the maximum linear conductance with second-order
perturbation theory for the whole range of temperatures. Surprisingly even
for such low-conductance SETs deviations from perturbation theory for
$G_{\text{min}}$ are pronounced at low temperatures. In contrast, a detailed
comparison with PIMC data for $g=4.75$ revealed good agreement between
experiment and theory outside the perturbative regime. Further comparison
showed that at $g=4.75$ not only $G_{\text{max}}$ and $G_{\text{min}}$ but also
the form of the Coulomb peaks could be described very well by our
simulations for temperatures $1/(\beta E_c) \ge 0.05$.

Finally we have presented a comparison of experimental and theoretical
results for the maximum and minimum conductance as a function of the
tunneling strength $g$ for a fixed temperature.  The experimental data for
this comparison are combined from experiments on different SETs including
also the results of earlier experiments by \citeauthor{joyez97}. 
For $g<8$ we find that the PIMC data is in good
accordance with experiment whereas perturbation theory in second order shows
significant deviations for $g>2.5$. Also the available data of third-order
perturbation theory still show rather large deviations for $G_{\text{max}}$.

The good agreement found in our study is an affirmation that the path
integral formulation in combination with the Monte Carlo method allows for
an accurate description of the minimum conductance $G_{\text{min}}$ over the
whole range of experimentally accessible parameters. Limitations of the
Monte Carlo method for the calculation of the maximum conductance
$G_{\text{max}}$ exist at low temperatures and small conductances due to slow
convergence.  Nonetheless $G_{\text{max}}$ as well as the entire shape of the
Coulomb peaks can be described well in a range of parameters which lies
outside the perturbative regime.

\begin{acknowledgments}
  We would like to thank Bernhard Obst (Forschungs\-zentrum Karlsruhe,
  Institut f\"{u}r Technische Physik) for supplying us with SEM images. We
  also thank B.\@ H\"{u}pper, P.\@ Joyez, J.\@ K\"{o}nig, H.\@ Schoeller,
  A.\@ I.\@ Yanson, and H.\@ v. L\"{o}hneysen for fruitful discussions.
  Financial support was provided by DFG through SFB 195 and SFB 276.  C.\@
  W.\@ and P.\@ v.\@ S.\@ acknowledge financial support by the
  Landesgraduiertenf\"{o}rderung Baden-W\"{u}rttemberg.
\end{acknowledgments}

\bibliography{set}

\end{document}